\documentclass[preprint,aps,amsmath,amssymb]{revtex4-2}

\newcommand{\be}{\begin{equation}}
\newcommand{\ee}{\end{equation}}
\newcommand{\bea}{\begin{eqnarray}}
\newcommand{\eea}{\end{eqnarray}}
\newcommand{\ba}{\begin{array}}
\newcommand{\ea}{\end{array}}

\usepackage{epsfig}
\usepackage{graphicx}
\usepackage{hyperref}
\usepackage{multirow}
\usepackage[utf8]{inputenc}
\usepackage{makecell}
\usepackage{natbib}

\begin{document}
\title{Exclusive diffractive $\mathrm{J/}\psi$  and $\psi(2S)$ production in dipole model using a holographic AdS/QCD light-front wavefunction with longitudinal confinement}
\author{Neetika Sharma}
\email{dr.neetika@nith.ac.in}
\affiliation{Department of Physics \& Photonics Science\, \\
National Institute of Technology, Hamirpur, HP-177005, India}

\begin{abstract} 
We use anti-de Sitter/quantum chromodynamics (AdS/QCD) based holographic light-front wavefunction  (LFWF) for vector meson, in conjunction with the  dipole model to investigate the cross sections data for  the diffractive and exclusive ${\mathrm J/}\psi$ and $\psi(2S)$ production. We confront the experimental data using a new  explicit form of the holographic LFWF,  where the longitudinal confinement dynamics in the light-front Schr\"odinger equation  is  captured by 't Hooft equation of (1 + 1)-dim, in large $N_c$ approximation, in addition to the transverse confinement dynamics governed by the confining mass scale parameter, $\kappa$ in vector mesons. We obtain the  LFWF parameters from fitting to the  exclusive
 $\mathrm{J/}\psi$ electroproduction data from electron-proton collision  at  the Hadron Electron Ring Accelerator for $m_c = 1.27 \mbox{ GeV}$. Our results suggest that the dipole model together with holographic meson LFWFs with longitudinal confinement is able to give a successful description for  differential  scattering cross section for  exclusive ${ \mathrm J/}\psi$ electroproduction for H1 and ZEUS data. We also predict the rapidity distributions of differential scattering cross section and total photoproduction of 
 ${\mathrm J/}\psi$ and $\psi(2S)$ states in proton-proton ultraperipheral collisions(UPCs) at center of mass energy $\sqrt s = 7, 13 ~\mbox{TeV}$. Using the minimum set of parameters, our predictions for the UPCs are in good agreement with the recent experimental observations of UPCs at ALICE and LHCb Collaborations.
\end{abstract}
\maketitle


\section{Introduction}
\label{Introduction}

Deep inelastic scattering (DIS) experiments and exclusive diffractive processes in electron-proton ($ep$) collisions, such as, deeply virtual Compton scattering (DVCS) and exclusive vector meson production (VMP), can provide valuable information about the parton saturation regime  at small-$x$ \cite{Nikolaev:1990ja}. The small $x$ physics is important because of their implications in our understanding of the parton density functions in ultra peripheral collisions (UPC) in proton-proton, proton-nuclei, and nuclei-nuclei interactions at the LHC, Relativistic Heavy Ion Collider and Large Hadron Electron Collider experiments. These days the high energy photons in UPCs  are extensively used as a probe to study the internal structure of protons and strong interaction dynamics in quantum chromodynamics (QCD). 

One of the thoroughly studied UPC processes at the LHCb is exclusive photoproduction of heavy vector mesons, in particular, photoproduction of $\mathrm{J/}\psi$ and $\psi(2S)$. Several measurements of exclusive $\mathrm{J/}\psi$ and $\psi(2S)$ production have been reported by the H1 and ZEUS Collaborations at the HERA $ep$ collider \cite{Aktas:2005xu,ZEUS:2004yeh}. The CDF Collaboration at the Tevatron collider reported the  exclusive charmonia production in $\mathrm{p}\mathrm{\bar{p}}$ collisions at $\sqrt{s} = 1.96 ~ \mbox{TeV}$ \cite{CDF:2009xey}. The first measurement of exclusive $\mathrm{J/}\psi$ and $\psi$(2S) production cross sections in ${\mathrm p}{\mathrm p}$ collisions was made by the LHCb  Collaboration at $\sqrt s = 7 ~ \mbox{TeV}$ \cite{LHCb:2013nqs,LHCb:2014acg} and  then at  $\sqrt s = 13 ~\mbox{TeV}$ \cite{LHCb:2018rcm}. These experimental measurements are significant as they extended the photon-proton center of mass energy to reach up to $ \sim 2 ~ \mbox{TeV}$.  Exclusive $\mathrm{J/}\psi$ photoproduction off protons in ultra peripheral ${\mathrm p}$-${\mathrm {Pb}}$ collisions at $\sqrt{s_{\rm NN}} = 5.02 ~\mbox{TeV}$ was investigated by the ALICE Collaboration   \cite{ALICE:2014eof,ALICE:2018oyo}. The ratio of cross sections  $\psi(2S) $ to ${\mathrm{J/}\psi}$ have been measured by the H1  \cite{ H1:2002yab}  and ZEUS Collaborations  \cite{ZEUS:2002wfj,ZEUS:2022sxn} for the exclusive and photoproduction data.  Further, more precise data on $\mathrm{J/}\psi$ photoproduction and exclusive production is expected from the  LHC run 3 to give better understanding of the parton distribution functions at small $x$.

The  dipole model is an important theoretical framework to provide an unified framework to study the exclusive and inclusive diffractive data in $ep$ collisions at HERA, and inclusive particle production data  in UPCs \cite{Nikolaev:1990ja,Mueller:1994jq}. An important ingredient for particle production in dipole model approach is that the scattering amplitude  for the singly diffractive processes $ \gamma^* p \to  V p$, where $V$ is a vector meson or a real photon $\gamma$ in DVCS, factorizes into a product of light-front wavefunctions (LFWFs) associated with the initial state photon and final state vector meson and a dipole-nucleon cross section. Though, the exact form of the photon wavefunction is known exactly in perturbative QCD, the vector meson wavefunction remains an open question.

In the recent past,  a new analytical insight into the meson LFWFs was proposed by  Brodsky and de T\'eramond in a semiclassical relativistic approximation in the light-front Schr\"odinger wave equation (LFSWE).  It has been shown that the LFSWE in the chiral limit  of massless quarks in physical space-time can be mapped with the equation of motion of spin-$J$ strings in the fifth dimension of anti-de Sitter ($\mathrm{AdS}$) space, thus establishing a connection between light-front QCD and $\mathrm{AdS_5}$, widely known as the Brodsky and de T\'eramond  ``Light-Front Holography'' (BdT-LFH) \cite{deTeramond:2005su,Brodsky:2006uqa,deTeramond:2008ht}. This approach has been successfully applied to investigate  the hadron spectroscopy, including the Regge trajectories,  hadron form factors, parton distributions,  infrared behavior of the strong coupling, etc. \cite{Brodsky:2013ar,Brodsky:2014yha,Sharma:2014voa}.  The precise mapping of  the electromagnetic/gravitational form factors in the $\mathrm{AdS}$ space with the corresponding expressions from light-front QCD in physical space-time leads to the exact form of holographic LFWFs, referred as the `BdT-LFWF'.  In  a series of previous work, it has been shown that the `BdT-LFWF'  together with the  dipole model  successfully predict the cross section for the diffractive $\rho$ \cite{Forshaw:2012im} and $\phi$  \cite{Ahmady:2016ujw}, when compared with the available data at the HERA  $ep$ collider \cite{Chekanov:2005cqa,Aaron:2009xp,Chekanov:2007zr}.

It is important  to  mention here that the  ``BdT-LFH'' approach discussed so far, addresses the chiral limit of massless quarks where only the transverse dynamics of hadrons were solved analytically and the  longitudinal light-front momentum fraction is frozen. Recently, there have been a lot of attention on incorporating the longitudinal  dynamics for understanding  the  chiral dynamics, contribution of non zero quark masses,  and  for identifying the physical states in the excited states spectrum, etc. One of the important  phenomenological models for the longitudinal confining interaction is based on one-gluon exchange interaction  in basis light-front quantization \cite{Weller:2021wog,Freese:2022ipu,Lyubovitskij:2022rod,Li:2022izo}. 

Recently, an another important approach viz. the 't Hooft model of (1+1)-dim, in large $N_c$ limit \cite{tHooft:1974pnl,*Zhitnitsky:1985um,*Grinstein:1997xk}, is employed successfully to go beyond the BdT prescription \cite{Chabysheva:2012fe}. It has been proven that the 't Hooft equation is complementary to, and consistent with, the LFHSE and provides a global description of the mass spectra of  ground and excited states of hadrons and tetraquarks, with a universal mass scale parameter, $\kappa$ \cite{Ahmady:2021lsh,Ahmady:2021yzh}. 

In view of the successes of 't Hooft approach, we  intend to use the LFWFs obtained by solving the 't Hooft equation in the large $N_c$ limit  to confront the data for  diffractive $\mathrm{J/}\psi$ and $\psi(2S)$ production with the HERA experiment. We will refer to this wavefunction as 't Hooft LFWF. For this purpose, we use the best fit set of parameters for the dipole model  \cite{Ahmady:2016ujw} obtained via fitting to the recent and precise HERA data on inclusive deep inelastic scattering \cite{Abramowicz:2015mha}. We obtain the parameters of the LFWFs via $\chi^2$ minimization to the recent measurement of the electroproduction data at HERA. Further, we intend to investigate the implication of the LFWFs in the exclusive $\mathrm{J/}\psi$ and $\psi(2S)$ photoproduction and rapidity distribution in proton-proton collisions in ultra peripheral as a probe of the QCD at high energies.

  
The plan of the work is as follows: We begin with briefly reviewing the dipole model in Sec. \ref{Dipole model} and then discuss the holographic 't Hooft light-front meson wavefunction in Sec. \ref{Holographic wfs}. We report the  predictions of the dipole cross section with the holographic meson wavefunction to compute the diffractive cross sections for $\mathrm{J/}\psi$  and $\psi(2S)$ production  and ultra peripheral collisions in Sec. \ref{Diffractive data}. We discuss results and conclusions in the  section  \ref{Conclusion}.

\section{The dipole model} 
\label{Dipole model}

It has been widely established that the exclusive diffractive VMPs and DVCS processes can be explained by the  dipole model of high-energy scattering.  In the dipole picture, the scattering amplitude for the diffractive process $\gamma^* p \to V p$ with an  exclusive final state vector meson $V \sim \rho, \phi, \mathrm{J/} \psi,\psi(2S), \Upsilon$ or a real photon $\gamma $ in DVCS in the forward limit, factorize into an overlap of  incoming photon and exclusive final state vector meson LFWF and a universal dipole-proton scattering  amplitude \cite{Iancu:2003ge}.  In the light-front form, the incoming virtual photon $\gamma^*$ at first fluctuates into a  neutral quark-antiquark pair ($q \bar q$) which forms a dipole. The  dipole  than interacts with the proton via gluon exchanges, and then the $q \bar q$ pair recombines to form either a final-state exclusive vector meson or a real photon. The lifetime of  a $q \bar q$ dipole at small $x$ is much larger than its typical interaction time with target proton or nuclei \cite{Watt:2007nr}.

The differential cross section for the exclusive VMP in the final state, can be expressed in terms of the imaginary part of scattering amplitude  \cite{Watt:2007nr,Rezaeian:2013tka}:
 \be
\left. {{\mathrm d} \sigma^{\gamma^*  p \to V p}_\lambda \over dt}\right. 
= \frac{1}{16\pi} 
| \Im \mbox{m}\, \mathcal{A}_\lambda(s,t;Q^2) |^2 \:.
\label{dsigmat0}
\ee
The scattering amplitude for the diffractive process in Eq. \eqref{dsigmat0}  can be expressed as:
\bea  \Im  \mbox{m}\, \mathcal{A}_\lambda(s,t;Q^2)  
 &=& 2 \sum_{h, \bar{h}} \int {\mathrm d}^2 {\mathbf r} \; \int {\mathrm d} x \; \int {\mathrm d}^2 {\mathbf b} \; \Psi^{\gamma^*,\lambda}_{h, \bar{h}}(r,x;Q^2)  \Psi^{V,\lambda}_{h, \bar{h}}(r,x) e^{-i[b- (1-x) \mathbf {r}] \cdot \mathbf{\Delta}} \mathcal{N}(x_{\text{m}},\mathrm{r}, \mathbf{b})\,, \nonumber \\
\label{amplitude-VMP}  \eea 
where $\mathcal{N}(x_{\text{m}},\mathrm{r},\mathbf{b})$ represents the proton-dipole scattering amplitude and $\mathbf{\Delta}^2 = - t$ is the squared momentum transfer at the proton vertex. The LFWFs for photon and proton $\Psi^{\gamma^*/V,\lambda}_{h, \bar{h}}(r,x)$ are the probability amplitudes for the virtual photon or vector meson to fluctuate into a $q\bar{q}$  dipole  for a given helicity configuration $h$ ($\bar {h}$) of the quark (antiquark). The LFWFs depend on the fraction of light-front momentum of the photon (or vector meson) carried by the quark $x$, the transverse separation between the quark and antiquark $r$, photon's virtuality $Q^2$, and impact parameter separation between the dipole and proton ${\mathbf b}$. The label $\lambda = L,T$  denotes the  photon or vector meson longitudinal and transverse polarizations. We will discuss in detail the explicit form of the LFWFs for the photon and mesons in the next section.

For the case of diffractive processes, the $b$-integration gives the four momentum transfer $t$ dependence of the cross section. In the forward limit \cite{Iancu:2003ge}, the impact parameter dependence of the dipole amplitude factorizes as: 
\be  \hat{\sigma}(q \bar q) = 2 \int {\mathrm d}^2 {\mathbf b}  \mathcal{N}(x_{\text{m}},\mathrm{r}, \mathbf{b}) = 2 \int {\mathrm d}^2 {\mathbf b}  \mathcal{N}(x_{\text{m}},\mathrm{r}) T(b)   = \sigma_0 \mathcal{N}(x_{\text{m}},\mathrm{r}) \label{sigma0}\,.\ee  In this way, the impact parameter dependence in Eq. \eqref{amplitude-VMP}  can be treated as an overall normalization factor $\sigma_0$. The factor $\sigma_0$ can be obtained via a fit to DIS data on $F_2$ structure functions. 

The expression for the scattering amplitude ${\mathcal N} (x,r)$  is obtained by smoothly interpolating between two limiting types of behavior for the varying dipole sizes. For small dipole sizes, $r \ll 2/Q_s$,  ${\mathcal N}$ is obtained from the saddle point approximation to the leading-order (LO) BK equation \cite{Balitsky:1995ub,Kovchegov:1999yj,Kovchegov:1999ua}, and for large dipole sizes, $r \gg 2/Q_s$, the Levin-Tuchin solution of the BK equation  inside the saturation region is used \cite{Levin:1999mw}. The dipole-proton scattering amplitude is therefore expressed as:
\bea { \mathcal N } (x_{\text{m}},r Q_s)  &=& { \mathcal N}_0 \left ( { r Q_s \over 2 }\right)^ {2 \left [ \gamma_s + { {\mathrm ln}  (2 / r Q_s) \over  k \, \lambda \, {\mathrm ln} (1/x_{\text{m}}) }\right]}  ~~~~~~~{\rm for } ~~~~~ ~~~~ r Q_s \leq 2 \nonumber \\ &=& { 1- \exp[-{\mathcal A} \,{\mathrm ln}^2 ( {\mathcal B} \, r Q_s)]}  ~~~~~~~~~~~{\rm for } ~~~~~ ~~~~ r Q_s > 2 \label{Nxr} \eea where the saturation scale  $Q_s \equiv  Q_s(x) = (x_0/x_{\text{m}})^{\lambda / 2}$ GeV. 
The coefficients ${\mathcal A}$  and ${\mathcal B}$ in Eq. \eqref{Nxr} are determined from the fact that the ${\mathcal N}(x_m, r Q_s)$ and its derivative with respect to $r Q_s$ are  continuous at $r Q_s=2$. This leads to the following expressions: \be  {\mathcal A} =  - { ({ \mathcal N}_0 \gamma_s)^2  \over (1 - { \mathcal N}_0)^2 \,  \ln [1 - { \mathcal N}_0] }\,, ~~~~~~ {\mathcal B} = {1 \over 2} (1 - { \mathcal N}_0)^{-{(1 - { \mathcal N}_0) \over { \mathcal N}_0 \gamma_s }}\,. \ee
The scattering amplitude for the elastic scattering of the dipole on the proton  contains all the high energy QCD dynamics of the dipole-proton interaction.  It depends on the photon-proton centre of mass energy $W$ via the modified Bjorken variable $x_{\mbox{m}}$ where $	x_{\text{m}}=x_{\text{Bj}}\left(1+ \frac{M_V^2}{Q^2} \right)~\text{with}~x_{\text{Bj}}=\frac{Q^2}{W^2} \;\,$  \cite{Rezaeian:2013tka}.

The dipole model parameter $k$ and  anomalous dimension coefficient $\gamma_s$ were fixed to 9.9, and 0.63 at the LO BFKL predictions  \cite{Iancu:2003ge}. The central fits are obtained at a fixed ${\mathcal N}_0$ = 0.7 and the other three parameters ($\sigma_0$, $x_0$ and, $\lambda$) were obtained via fitting the $F_2$ structure functions  data \cite{Watt:2007nr}.  Further, it has been shown \cite{Soyez:2007kg} that allowing the anomalous dimension parameter at the saturation scale to vary to a higher value $\gamma_s$ = 0.74, improves the fit to the $F_2$ data. This predicted value of $\gamma_s$ is also significantly close to the value extracted from the  renormalisation-group-improved next-to-leading order  BFKL kernel calculations \cite{Salam:1998tj}.

Following Ref. \cite{Rezaeian:2012ji},  choosing the profile function $T(b)$ in Eq. \eqref{sigma0} in the form  \be T(b)= {1 \over 2 \pi  B_D} \exp \left[- {b^2  \over 2 B_D} \right]\,, \label{tb} \ee  where $B_D$ is the diffractive slope parameter. Substituting Eq. \eqref{tb} in the Fourier transform over $b$ in Eq. \eqref{amplitude-VMP}, gives the diffractive cross section decreasing exponentially with $t$ as $\exp(-B_D |t|)$ \cite{Golec-Biernat:1999qor,ZEUS:2002wfj,H1:2005dtp,ZEUS:2022sxn}. 
The $t$ dependence of the cross section in exponential form is  expressed as:
\be
\left. {{\mathrm d} \sigma_\lambda \over dt}\right.
= \frac{1}{16\pi} [\mathcal{A}_\lambda(s, t=0)]^2  \exp(- B_D|t|)\,,
\label{dsigmadt}
\ee 
where $B_D$ is expressed as \be B_D = N\left( 14.0 \left(\frac{1~\mathrm{GeV}^2}{Q^2 + M_{V}^2}\right)^{0.3}+1\right)\,, \label{Bslope} \ee with $N=0.55$ GeV$^{-2}$ in accordance with the experimental data \cite{Caldwell:2001ky,Forshaw:2006np}.


The contribution of the real part of the scattering amplitude can be incorporated in  Eq. \eqref{dsigmadt}  for the differential scattering amplitude as follows:
\be
\left. {{\mathrm d} \sigma_\lambda \over dt}\right.
= \frac{1}{16\pi} 
[\Im \mbox{m}\, \mathcal{A}_\lambda(s,t;Q^2) ]^2 \;  \exp(-B_D |t|) (1 + \beta_\lambda^2)
\label{dsigmadt-re}
\ee
where the factor $\beta_\lambda$ is the ratio of real to imaginary parts of the scattering amplitude expressed as \cite{Kowalski:2006hc} :
\be \beta_\lambda=\tan \left(\frac{\pi}{2} \alpha_\lambda \right) ~~ ~  {\text{with} }~~ ~  \alpha_\lambda=\frac{\partial \ln | \Im \mbox{m}\, \mathcal{A}_\lambda(x,t)|}{\partial \ln  \left(1/x\right)} \label{logderivative} \,. \ee 

Integrating differential scattering cross section in Eq. \eqref{dsigmadt-re} over $t$, one can obtain the transverse and longitudinal contribution to the cross section. We considered the total cross section for  $\mathrm{J/}\psi$ electroproduction is: $ \sigma = \sigma_T + 0.98\, \sigma_L$ to be compared with the experimental data.

We will  now briefly discuss the photon and vector meson light-front wavefunction.  The explicit form of the photon light-front wavefunction is obtained in the perturbative methods in light-front QED in $r$ space. The photon wavefunction  to the lowest order in the electromagnetic coupling $\alpha_{\mathrm{EM}}$ is expressed as \cite{Lepage:1980fj,Dosch:1996ss,Forshaw:2003ki,Kulzinger:1998hw}:
\bea \Psi^{\gamma,L}_{h,\bar{h}}(r,x; Q^2, m_f)  &=& \sqrt{\frac{N_{c}}{4\pi}}\delta_{h,-\bar{h}}e\, e_{f}2 x(1-x) Q \frac{K_{0}(\epsilon r)}{2\pi}\;, \label{photonwfL} \\ 
\Psi^{\gamma,T}_{h,\bar{h}}(r,x; Q^2, m_f) &=& \pm \sqrt{\frac{N_{c}}{2\pi}} e \, e_{f}   \big[i e^{ \pm i\theta_{r}} (x \delta_{h\pm,\bar{h}\mp} -  (1-x) \delta_{h\mp,\bar{h}\pm}) \partial_{r}   +  m_{f} \delta_{h\pm,\bar{h}\pm} \big]\frac{K_{0}(\epsilon r)}{2\pi} \,, \label{photonwfT} \\ \nonumber \eea
where $ \epsilon^{2} = x(1-x)Q^{2} + m_{f}^{2} $ and $K_0(\epsilon r)$ is the modified Bessel function.  Here the quark mass $m_f$ acts as an infrared regulator as at $Q^2 \to 0$ or $x \to (0,1)$, and the photon light-front wavefunctions become sensitive to the nonzero quark mass $m_f$, which prevents the modified Bessel function $K_0(\epsilon r)$ from diverging.

We will now discuss the vector mesons light-front wavefunction, however, the explicit form of the wavefunction is not completely known using the present perturbative techniques.  In this work, we  follow the nonperturbative prescription in Ref. \cite{Forshaw:2003ki} for predicting the meson light-front wavefunction.  Taking an analogy from the photon LFWF, the helicity dependent vector meson  LFWF consisting of a spinor, scalar part, and an unknown nonperturbative wavefunction are expressed as \cite{Forshaw:2012im, Ahmady:2016ujw}:
 \be \Psi^{V,L}_{h,\bar{h}}(r,x) =  \frac{1}{2} \delta_{h,-\bar{h}}  \bigg[ 1 +  { m_{f}^{2} -  \nabla_r^{2}  \over x(1-x)M^2_{V} } \bigg] \Psi_L(r,x) \,,\label{MesonL} \ee and \be \Psi^{V, T}_{h,\bar{h}}(r,x) = \pm \bigg[  i e^{\pm i\theta_{r}}  ( x \delta_{h\pm,\bar{h}\mp} - (1-x)  \delta_{h\mp,\bar{h}\pm})  \partial_{r}+ m_{f}\delta_{h\pm,\bar{h}\pm} \bigg] {\Psi_T(r,x) \over 2 x (1-x)}\,,\label{MesonT} \ee
where $\Psi_{h, \bar h}(r,x)$ is the non-perturbative part of the wavefunction.

There are various prescriptions for modeling the nonperturbative $\Psi_h(r,x)$ part of the meson wavefunction.  Variants of quark models indicate that hadrons at rest can be modeled simply by considering a Gaussian fluctuations in transverse space, referred to as the  boosted Gaussian (BG) wavefunction  \cite{Dosch:1996ss,Nemchik:1996cw, Forshaw:2003ki}.  BG wavefunction is self-consistent, fully boost-invariant, and has a proper short-distance  behaviour of $ x(1-x)$ in the limit of massless quarks. It has been used widely  to describe simultaneously the cross section data on diffractive vector meson $\rho, \phi$, and $\mathrm{J/}\Psi$ production using the CGC,  impact parameter dependent  models, and IP-sat models  \cite{Forshaw:2010py,Forshaw:2011yj, Rezaeian:2012ji,Rezaeian:2013tka}. Recently there has been interest in using the dipole cross section extracted in Ref. \cite{Rezaeian:2013tka} with the same BG wavefunction to predict vector meson production in ultrapheripheral collisions at the LHC  \cite{Santos:2014vwa}.

 \section{ Longitudinal confinement in Holographic  meson  wavefunctions }
\label{Holographic wfs}

In the past decade, a new insight into the meson holographic light-front wavefunctions $ {\Psi_{L,T}(r,x)}$ have been obtained based on the semiclassical approximation of light-front QCD with massless quarks, referred as the  LFH \cite{deTeramond:2005su,Brodsky:2006uqa, deTeramond:2008ht}.  This approach, pioneered by Brodsky and de T\'ramond, determines the effective potential based on the unique mapping between the equation of motion of the string spin $J$ modes in the AdS space  and the  Hamiltonian formulation of (3 + 1)-dim QCD on the light-front form, to obtain a Schr\"odinger-like equation for the hadrons. The semiclassical approximation is  successful in explaining hadron spectroscopy, including tetraquarks and exotic states, form factors,  parton distribution, and the behavior of the QCD running coupling in the nonperturbative domain, etc.

The valence meson light-front wavefunction in light-front QCD are expressed as \cite{Brodsky:2014yha}  \begin{equation} \Psi(\zeta, x, \varphi)=  e^{iL\varphi} \frac{\phi (\zeta)}{\sqrt{2 \pi \zeta}}  {\it{X}}(x) \,,  \label{mesonwf} \end{equation} where $L \equiv |L_{z}^{max}|$  being the light-front orbital angular momentum quantum number, $\it{X}(x)$ is the  longitudinal, and $\phi(\zeta)$ is the transverse part of wavefunction. In order to completely specify the holographic wavefunction given by Eq. \eqref{mesonwf}, we need to specify the longitudinal  ${\it{X}}(x)$ and transverse part  $\phi(\zeta)$ of wavefunction. In ``BdT approach'', the longitudinal modes ${\it{X}}(x)  = \sqrt{x(1-x)} \chi(x) $  are fixed from the mapping of electromagnetic (or gravitational) form factors in physical space time to the $\mathrm{AdS}$ space \cite{Brodsky:2014yha}. The holographic mapping leads to $\chi(x)=1$,  giving the solution of longitudinal modes: $ {\it X}(x) = \sqrt{x(1-x)} \,.$ 


The transverse wavefunction $\phi(\zeta)$ is the solution of the  following LFHSE: \begin{equation} \left(-\frac{d^2}{d\zeta^2} + \frac{4L^2+1}{4\zeta^2} + U_{\perp}(\zeta) \right) \phi(\zeta) = M^2 \phi(\zeta) \,, \label{holograhicSE} \end{equation} where  $U_{\perp}(\zeta)$ is the confining potential at equal light-front time, and $M$ is the meson mass. The form of confining potentially is uniquely determined  from  the underlying conformal symmetry and a holographic mapping of  variable $\zeta $ to  the fifth dimension of $\mathrm{AdS}$ space gives $\zeta = {\sqrt{x(1-x)}} {\bf r}_\perp $.  A quadratic dilaton field  $ (\varphi(z) =\kappa^2 z^2)$ in the AdS background, breaks the conformal invariance and leads to  a harmonic oscillator potential in physical space-time as \cite{Brodsky:2013npa}: \begin{equation} U_\perp(\zeta, J)= \kappa^4 \zeta^2 + \kappa^2 (J-1) \;. 	\label{harmonic-LF} \end{equation}

Solving the LFHSE with the harmonic potential given by Eq. \eqref{harmonic-LF}  leads to the normalized eigenfunctions  for meson: \begin{equation} \phi_{n,L}(\zeta)= \kappa^{1+L} \sqrt{\frac{2 n !}{(n+L)!}} \zeta^{1/2+L} \exp\left(\frac{-\kappa^2 \zeta^2}{2}\right) L_n^L(x^2 \zeta^2) \;. \label{phi-zeta}  \end{equation}  
Importantly,  this prescription  predicts the massless pions as expected in the chiral limit. It also correctly predicts the Regge-like linear dependence of the meson mass squared  on the radial $n$ and orbital quantum numbers $L$. This information is further used to constraint the parameter $\kappa$ from  the fit to the Regge slopes data \cite{Workman:2022ynf}. Reference \cite{Brodsky:2014yha} reports a universal $\kappa \sim 0.54$ GeV for vector mesons.
 
The complete  wavefunction for mesons in the position space  with the quark mass terms  and polarizing part   can be expressed as: \be  \Psi^{\lambda}_{\rm BdT} (r,x) = \mathcal{N}_{\lambda}\sqrt{x (1-x)}  \exp \left[ -{ \kappa^2  x (1-x) r^2_{\perp} \over 2} \right] \exp \left[ -{m_f^2 \over 2 \kappa^2 x(1-x) } \right] \,, \label{hwf1} \ee
 where  the polarization-dependent normalization constant ${\mathcal N}_{\lambda}$ to be calculated using the following normalization condition:  \be \sum_{h,\bar{h}} \int {\mathrm d}^2 {\mathbf{r}} \, {\mathrm d} x | \Psi^{V, \lambda} _{h, {\bar h}}(r,x)|^{2} = 1 \,. \label{normalisation}  \ee

Recently, there has been a lot of interest in understanding the longitudinal wavefunction ${\it{X}}(x)$ dynamically for performing the more realistic calculations of the mass spectrum of hadrons.  The main idea is to separate out the light-front Scho\"rdinger equation for the longitudinal part that includes a dynamical confining potential along with the contribution of quark mass. Using this approach, a  new phenomenological longitudinal potential in the basis light-front quantization approach was proposed $ U_\parallel^{\mathrm{BLFQ}}(x)=-\sigma^2 \partial_x (x(1-x)) \partial_x\,, $ where $\sigma$ is a mass scale parameter  to explain the ground states of light and heavy mesons including their excited states \cite{Li:2015zda,Li:2021jqb,DeTeramond:2021jnn}. 

Another class of longitudinal potential models based on instantaneous gluon exchange potential are the ones obtained from the 't Hooft model in the large-$N_c$ limit \cite{tHooft:1974pnl,Grinstein:1997xk}. The 't Hooft model based potential is confining in nature and yields a longitudinal wave function consistent with the  $\mathrm{AdS/QCD}$ duality in the  chiral limit. A few decades ago, the 't Hooft model  was used extensively to investigate the properties of mesons, such as, confinement, Regge trajectories, etc. \cite{ Zhitnitsky:1985um, Grinstein:1992ub, Grinstein:1997xk, Ma:2021yqx}. Recently, the idea of exploiting the 't Hooft model as a possible way out  to go beyond the BdT prescription and add longitudinal confinement was first proposed in Ref. \cite{Chabysheva:2012fe} for  predicting the decay constant of mesons. In  a series of  recent  work \cite{Ahmady:2021lsh,Ahmady:2021yzh}, authors have solved  the 't Hooft equation following the work of Ref. \cite{Chabysheva:2012fe}, together with the holographic Schr\"odinger wquation, to provide a global description of the mass spectrum of the ground and excited states of mesons, baryons, and tetraquark with a universal mass scale $\kappa$. These recent developments in the 't Hooft model  make it  a promising avenue for confronting  the diffractive $\mathrm{J/}\psi$ and $\psi(2S)$ vector mesons  data.

We will now briefly discuss the 't Hooft formalism. Note that  't Hooft derived a Schr\"odinger-like equation for the longitudinal modes, starting  from the QCD Lagrangian in $(1+1)$-dim and in the large $N_c$ approximation  \cite{tHooft:1974pnl}: \begin{equation}\left(\frac{m_f^2}{x}+\frac{m_{\bar{f}}^2}{1-x}\right)X(x) + U_L(x) X(x)=M^2_L X(x) \;,   \label{thoofteq} \end{equation} with \be 	U_L(x)X(x)=\frac{g^2}{\pi} \mathcal{P} \int {\rm d}y \frac{|X(x)-X(y)|}{(x-y)}^2 \,, \label{thooft-potential} \ee where $g \equiv g_s \sqrt{N_c}$ is the finite 't Hooft coupling constant with mass dimensions and $\mathcal{P}$ denotes the principal value prescription. 

It is worth mentioning here that while the holographic Schr\"odinger equation is solved analytically, the 't Hooft equation does not have any exact analytical solution. To find the numerical solution of the 't Hooft equation \eqref{thoofteq},  we expand the longitudinal mode onto a Jacobi polynomial basis $\chi(x) = \sum_n C_n f_n$, where $C_n$ are the  expansion coefficient and $f_n$ is the Jacobi polynomial basis \cite{Chabysheva:2012fe}. The resulting matrix representation of Eq. \eqref{thoofteq} is then diagonalized numerically in $f_n$ basis to find the wavefunction. This procedure guarantees that predicted masses are independent of the choice of basis and remain stable with respect to variations of basis parameters.  For equal-mass cases, the longitudinal equation obeys an $ x \leftrightarrow 1-x $ symmetry and the  resultant 't Hooft wavefunction for  mesonic states can be expressed as:
\be  \Psi^{\lambda}_{\rm 't ~ Hooft} (r, x) \sim x^{\beta}{(1-x)}^{\beta}   \exp \left[ -{ \kappa^2  x (1-x) r^2_{\perp} \over 2} \right]\,. \label{hwf1} \ee 


\section{Predicting diffractive cross sections and ultra peripheral proton-proton Collisions}

\label{Diffractive data}

Having specified the formalism of dipole cross section and the 't Hooft LFWF, we will now discuss the various parameters, needed to compute cross sections for diffractive $\mathrm{J/}\psi$  and $\psi(2S)$ production.   We use the  dipole model parameter $k = 9.9$ (LO BFKL values) and  ${\mathcal N}_0 =0.7$ (central fit) \cite{Iancu:2003ge}. Other model parameters $\sigma_0, \lambda, x_0$, and $\gamma_s$  were fixed  from a high quality DIS data  on structure function $F_2$ and reduced cross sections from HERA \cite{Watt:2007nr}.  
The idea behind the fitting is that the dipole-proton scattering amplitude  is a universal object and appear in the formula for the fully inclusive DIS process: $\gamma^* p \to  X$. If we replace the vector meson in the final state by a virtual photon in Eq. \eqref{amplitude-VMP}, one can obtain the cross section for DIS \cite{Forshaw:2006np, Kowalski:2006hc, Marquet:2007qa, Watt:2007nr, Rezaeian:2012ji, Rezaeian:2013tka}.

Using this procedure,  the recent extraction of the dipole model parameters was performed in Ref. \cite{Ahmady:2016ujw} using the highly precise combined data  for the reduced cross section given by  the H1 and ZEUS Collaborations \cite{Abramowicz:2015mha}. The best fit set for the dipole model parameters  together with the resulting  $\chi^2$ per degrees of freedom ($\chi^2/\mbox{d.o.f}$) are presenting in Table \ref{tab:F2fit}. The same set of parameters have been successfully used for confronting the data for  diffractive $\rho$ and $\phi$ production in Ref.  \cite{Ahmady:2016ujw}, thereby testing the success of dipole model parameters to the holographic wavefunction for the light vector mesons.  In this work, we use the same set of parameters  to  make predictions for the diffractive $\mathrm{J/}\psi$  and $\psi(2S)$ production. 

\begin{table}[h]
  \centering
  \begin{tabular}{c|c|c|c|c|c|c|}
    \hline\hline
      $[m_{u,d},m_s]$GeV  & $[m_c]$ GeV  &$\lambda$ & $\gamma_s$ ~~  &~~$\sigma_0$/mb ~~& $x_0$ &  $\chi^2/\mbox{d.o.f}$ \\ \hline
 $ [0.14,0.14]$ &1.27 & 0.206 &0.724 &29.9  &  $6.33 \times 10^{-6}$ &1.07 \\ \hline
       \end{tabular}
  \caption{Best fit parameters of the dipole model from $\chi^2$ fit to inclusive DIS data (with $x_{\text{Bj}} \le 0.01$ and $Q^2\in [0.045, 45]\,\text{GeV}^2$)  for the charm quark mass $m_c=1.27~ \mbox{GeV}$. }
  \label{tab:F2fit}
\end{table}

In addition to the dipole model parameters, we find the best fit set of parameters for the 't Hooft LFWF to investigate the  $\mathrm{J/}\psi$ and $\psi(2S)$ production diffractive HERA data.  For the sake of generalization of the LFWF, we  consider a parametrization that accommodates the numerical solution of 't Hooft LFWF in Eq. \eqref{thoofteq}  and is very much similar to the  Boosted Gaussian form of LFWF. The complete wavefunction after adding the polarization part is as follows:
\be  \Psi^{ \mathrm{J/}\psi, \lambda}_{\rm 't~ Hooft} (r, x) = \mathcal{N}_{\lambda}  { x^{\beta} (1-x)^{\beta}} \exp \left[ -{ \kappa^2  x (1-x) r^2_{\perp} \over 2} \right]  \,, \label{twf1} \ee
Similarly for $\psi(2S)$, we have  \be  \Psi^{ \psi(2S), \lambda}_{{\rm 't~ Hooft}} (r, x) = \mathcal{N}_{\lambda}  { x^{\beta} (1-x)^{\beta}} ( 1- \alpha \kappa^2 x(1-x) r^2_{\perp}  ) \exp \left[ -{ \kappa^2  x (1-x) r^2_{\perp} \over 2} \right]  \,. \label{twf2} \ee where  ${\mathcal N}_{\lambda}$ is the polarization-dependent normalization constant with the following condition:  \be \sum_{h,\bar{h}} \int {\mathrm d}^2 {\mathbf{r}} \, {\mathrm d} x | \Psi^{V, \lambda} _{h, {\bar h}}(r,x)|^{2} = 1 \,. \label{normalisation}  \ee

It is customary to obtain important constraints on the LFWF parameters of $\mathrm{J/}\psi$ and $\psi(2S)$ from the experimental data on electronic decay width and electroproduction data. We start by computing the chi-square per data point ($\chi^2/\text{d.p.}$)  in the ($\beta, \kappa$) parameter space for the available electroproduction data from HERA (56 data points)  \cite{Aktas:2005xu,ZEUS:2004yeh} and electronic decay width \cite{Workman:2022ynf} for $\mathrm{J/}\psi$ state. Our predicted best fit has a $\chi^2/\text{d.p.} = 34/57 = 0.60$ and is achieved with $\kappa = 1.41 ~ \mbox{GeV}$ and $\beta$ = 4.62.  The new parameter $\alpha$ in $\psi(2S)$ wavefunction is constrained from the orthogonality conditions for the mesonic  states.   Further, better fits to the data are even possible, if we allow the parameters $\beta$ and $\kappa$ to depend on the polarization of the  $\mathrm{J/}\psi$ and $\psi(2S)$  but that will add  additional free parameters in the wavefunctions.  On the other hand, the numerical solution of Eq. \eqref{thoofteq} giving a good fit  simultaneously to the $\mathrm{J/}\psi$ and $\psi(2S)$ states mass spectrum and Regge slopes is obtained using  $g = 0.523$ and $m_c = 1.370$, and predicts: $\beta = 4.5$. This result also confirms that numerical solution of 't Hooft LFWF ($\beta = 4.5$) is significantly close to the our minimized solution ($\beta = 4.62$).  It is also important to mention here that our minimization procedure also restricts the value $\kappa = 1.4 \pm 0.1$ and ignore the idea of universal $\kappa$ for the mesons diffractive data.

Having specified the parameters for the  dipole model  in Table \ref{tab:F2fit} and best fit for 't Hooft LFWF ($\kappa = 1.41 ~\mbox{GeV}$, $\beta = 4.62$, and $\alpha=0.74$), we now compute the cross sections for diffractive $\mathrm{J/}\psi$ and $\psi(2S)$ production. Recall that our prediction are with a new holographic light-front wavefunction specified by Eqs. \eqref{twf1} $\mathrm{J/}\psi$  and \eqref{twf2} for $\psi(2S)$ vector meson, we shall refers to these prediction as the 't Hooft LFWF (black solid curves). For $\mathrm{J/}\psi$, we predict the total cross section as a function of $Q^2$  as well as  $W$ in different $Q^2$ bins. We  predict the ratio of longitudinal to transverse cross sections as a function of $Q^2$ at fixed $W$. We also report the differential scattering cross section  given by Eq. \eqref{dsigmadt-re} as a function of $t$ at fixed $W$ for H1 and ZEUS Collaboration.

\begin{figure}[htbp]
\centering 
\includegraphics[width=8cm,height=8cm]{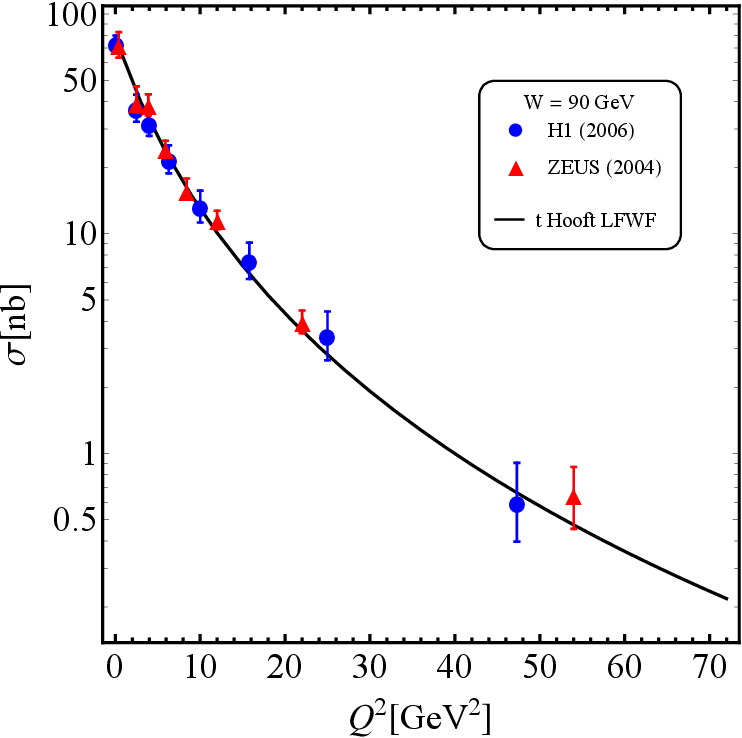}
\caption{Predictions for $\mathrm{J/}\psi$ electroproduction cross section as a function of $Q^2$ at fixed $W=90 \mbox{GeV}$, compared to H1 and ZEUS data \cite{Aktas:2005xu,ZEUS:2004yeh}. }
\label{jpsi:sigmaQ2}
\end{figure}

\begin{figure}[htbp]
\centering 
\includegraphics[width=8cm,height=9cm]{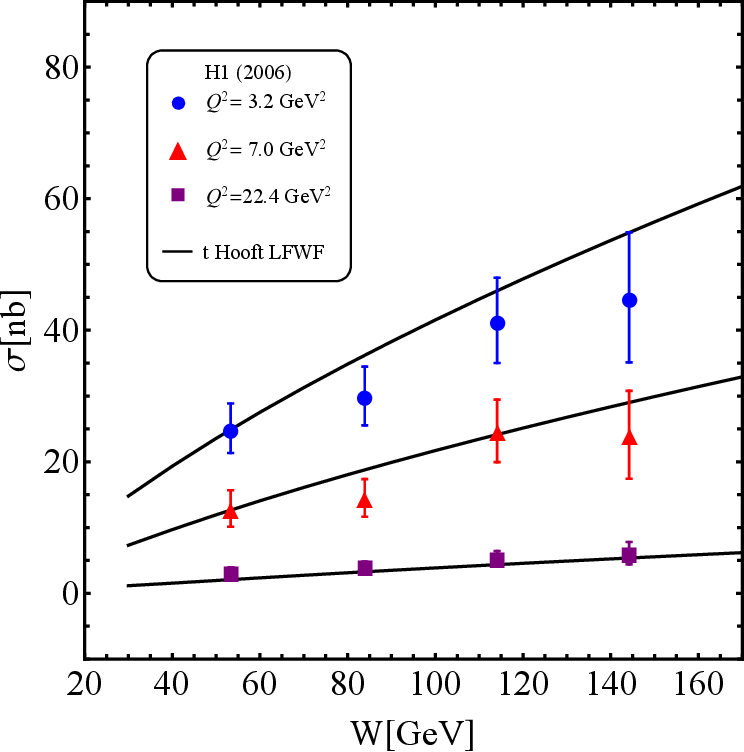}
\includegraphics[width=8cm,height=9cm]{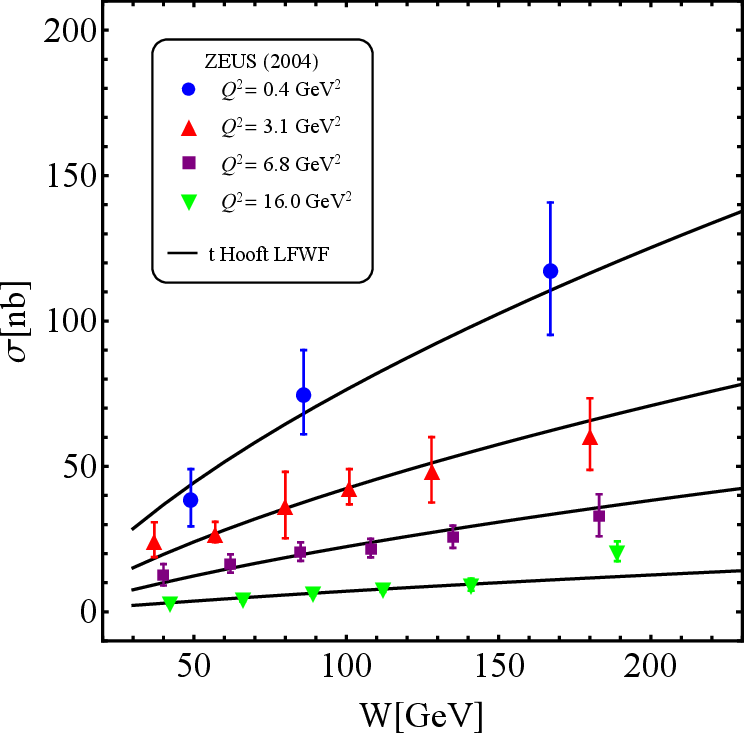}
\caption{Predictions for $\mathrm{J/}\psi$ electroproduction cross section $\sigma$ as a function of $W$ in different $Q^2$ bins compared to H1 and ZEUS data \cite{Aktas:2005xu,ZEUS:2004yeh}. }
\label{jpsi:sigmaW-H1-ZEUS}
\end{figure}

\begin{figure}[htbp]
\centering 
\includegraphics[width=10cm,height=10cm]{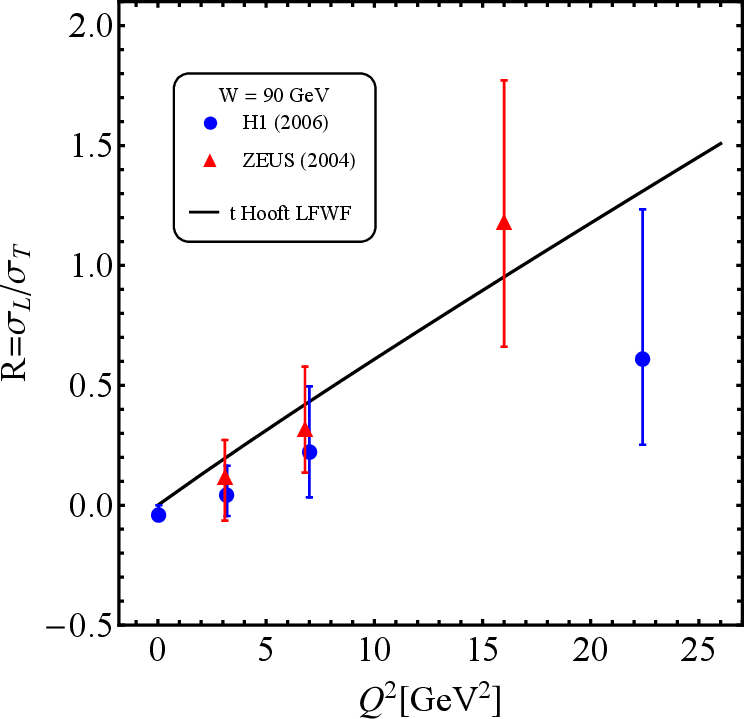}
\caption{Predictions for the longitudinal to transverse cross section ratio in $\mathrm{J/}\psi$ production as a function of $Q^2$ at $W=90 ~\mbox{GeV}$ compared to the H1 and ZEUS data  \cite{Aktas:2005xu,ZEUS:2004yeh}. }\label{jpsi:Ratio}
\end{figure}

\begin{figure}[htbp]\centering 
\includegraphics[width=8cm,height=8cm]{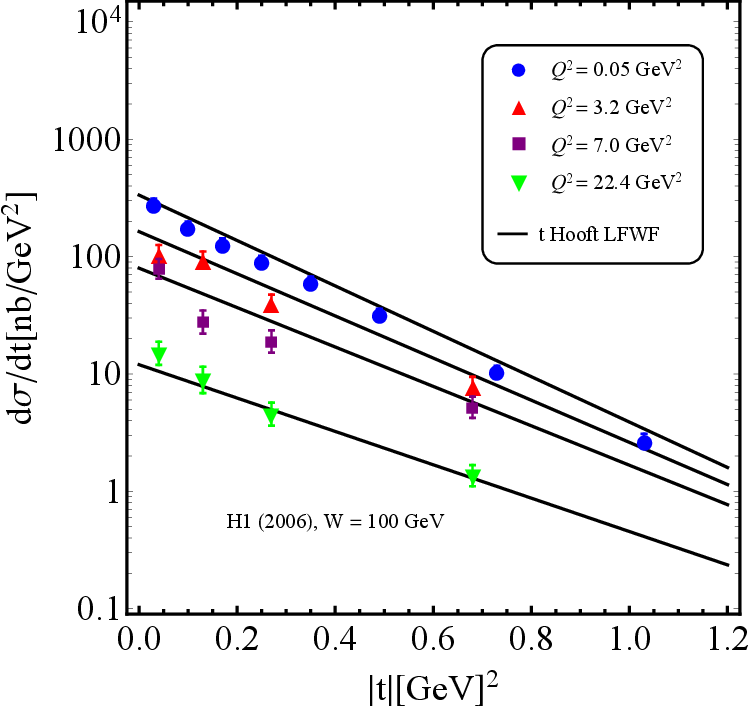}
\includegraphics[width=8cm,height=8cm]{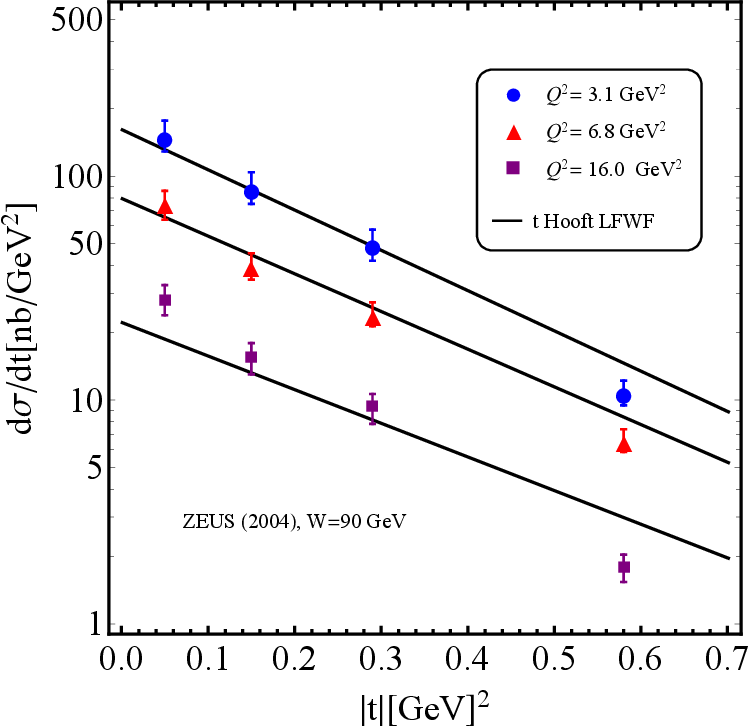}
\caption{\small Predictions for differential scattering cross section for $\mathrm{J/}\psi$ electroproduction as a function of $t = - \bf \Delta^2$ in different $Q^2$ bins compared to H1 and ZEUS Collaboration \cite{Aktas:2005xu,ZEUS:2004yeh}. }
\label{jpsi:dsigmavst}
\end{figure}

For $\mathrm{J/}\psi$ production, our predictions for the $Q^2$ dependence of the total cross section at fixed $W = 90 ~ \mbox{GeV}$  are shown in Figure \ref{jpsi:sigmaQ2} while  $W$ dependence of the total cross section in different $Q^2$ bins are shown in Figures \ref{jpsi:sigmaW-H1-ZEUS}. We confronted our prediction with the H1 and ZEUS data \cite{Aktas:2005xu,ZEUS:2004yeh} for $\mathrm{J/}\psi$ production in various kinematics. The  ratio of longitudinal to the transverse cross section $(\sigma_L/\sigma_T)$ data, shown in Figure \ref{jpsi:Ratio}.   Predicting the ratio is interesting since the normalization uncertainties in the diffractive $B_D$ slope and the dipole cross section, cancel out, increasing its sensitivity to the fitted parameters of meson wavefunction. Notice that our fitted parameter give a  fit over a wider range of $Q^2$. We also predicted the variation of the differential scattering cross section  with $t ~[\mbox{GeV}^2]$ at fixed $W = 100 (90) ~\mbox{GeV}$ in Figure \ref{jpsi:dsigmavst} for the H1 and ZEUS Collaboration. It is generally seen that the agreement between our results and data is excellent for the charm mass $m_c = 1.27 \mbox{GeV}$.

\begin{figure}[htbp]
\centering 
\includegraphics[width=9cm,height=9cm]{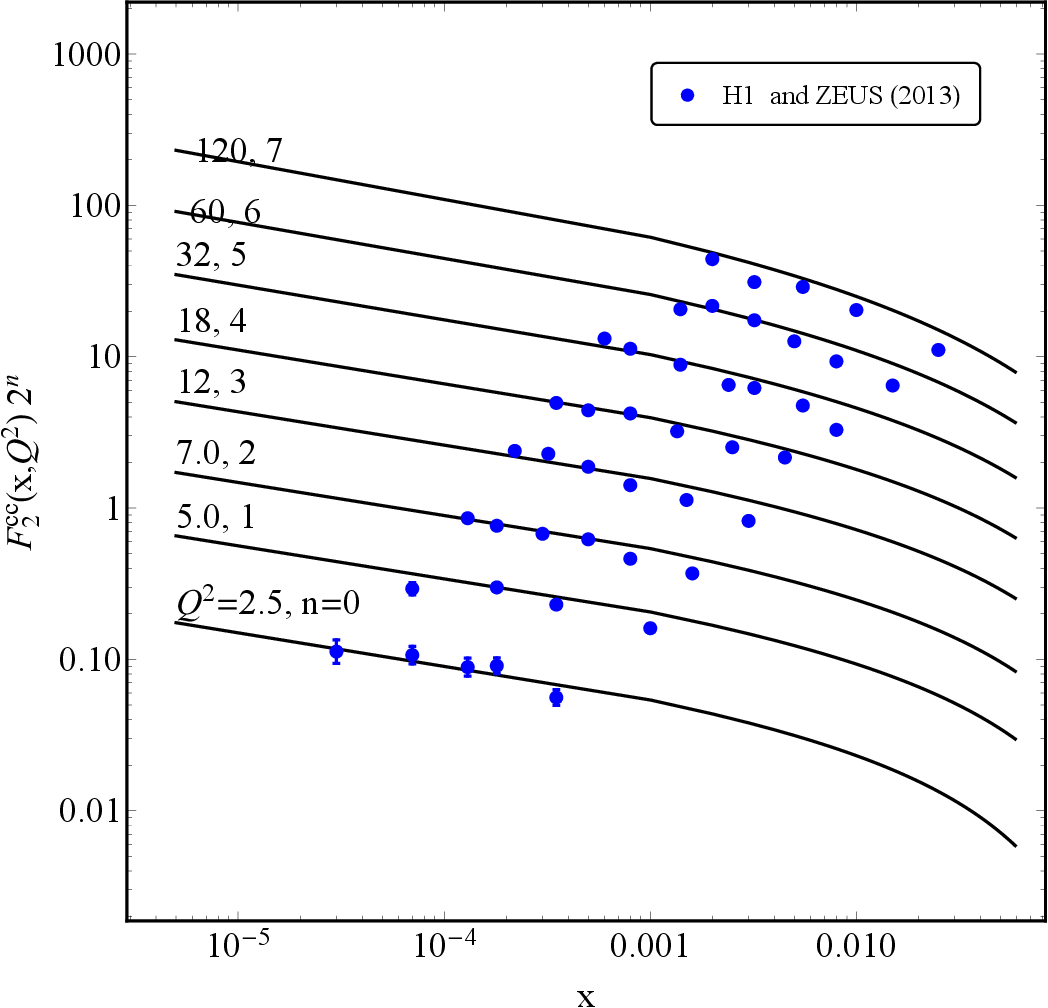}
\caption{Predictions for the combined Proton charmed structure function $F^2_{c \bar c} \sim  \sigma_r^{c \bar c}$ as a function of $x$ in different $Q^2$ bins. The experimental data points are from the combined data set from the H1 and ZEUS Collaboration \cite{Lipka:2013yen}.  }
\label{jpsi:f2cc}
\end{figure}

For a non-trivial consistency check of the dipole model parameters, we have also given a comparison with the structure functions $F_2^{c\bar c}$ in Fig. \ref{jpsi:f2cc}. We have predicted the  $x$ variation of $F_2^{c\bar c}$ and compared with H1 and ZEUS combined analysis data assuming that the contribution of $ F_2^{c\bar c}$ is close to the reduced cross section \cite{Forshaw:2006np}. The agreement of dipole model predictions for the structure function with the experimental data is striking upto $x < 0.01$ as the parameters of the dipole model are  fitted with $x<0.01$ and thus agreement begins to fail closer to $x = 0.01$. We have extended our model to the  range beyond the kinematics of existing data, as a prediction for the  future DIS experiments.

Now we will study the rapidity distribution for the  vector mesons production in  the ultra peripheral  collisions (UPC) in  the process $\mathrm{p}  \mathrm{p} \to \mathrm{p} \otimes V \otimes \mathrm{p}$, where the sign $\otimes$ denote rapidity gaps and $V$ stands for $\mathrm{J/}\psi $  and $\psi(2S)$  mesons. The measurement of UPCs are important as they allow a probe of the gluon distribution function at low values of $x$. Recently, the LHCb Collaboration have reported preliminary measurements of exclusive $\mathrm{J/}\psi $  and $\psi(2S)$ photoproduction in UPC  at $ 13 ~\mbox{TeV}$ \cite{LHCb:2018rcm}. More refined photoproduction data is expected soon from LHC run 3. In view of these developments, we intend to extend the dipole model with 't Hooft LFWF to investigate the rapidity distribution of exclusive $\mathrm{J/}\psi$ and $\psi(2S)$ production in UPCs. 
 
 Theoretically, the rapidity distributions of heavy meson production in UPCs is the product of photoproduction cross sections $\sigma^{\gamma\mathrm{p}\to V \mathrm{p}}$, the photon flux factor, and a rapidity gap survival factor. The differential cross section for the exclusive $\mathrm{J/}\psi$ photoproduction off proton in ${\mathrm p}{\mathrm p}$ UPCs  
is expressed as \cite{Jones:2013pga,GayDucati:2013sss}
\begin{eqnarray}
\frac{ {\mathrm  d} \sigma [{ \mathrm{p}  \mathrm{p} \to \mathrm{p} \otimes J/\psi \otimes  \mathrm{p}  }]}{d \mathrm{y}}  = S^2(W^+) \left(k^+ \frac{\mathrm{d} \mathrm{n}}{\mathrm{d} \mathrm{k^+} }\right) \sigma^{\gamma p\to \mathrm{J/}\psi p}(W^+)+ S^2(W^-) \left(k^-\frac{\mathrm{d} \mathrm{n} }{\mathrm{d} \mathrm{k^-}} \right) \sigma^{\gamma p\to \mathrm{J/}\psi p}(W^-)\,,\nonumber \\
\label{dsdy}
\end{eqnarray}
where $k^\pm$ is momentum of the photon radiated from proton, related to the rapidity $\mathrm{y}$ of the vector meson in the final state via the relation: $k^{\pm}={M_V \over 2} \exp(\pm |\mathrm{y}|)$.  In UPC, $\mathrm{W}$ is the center of mass energy of the photon-proton system, $W^{\pm}=(2k^\pm\sqrt{s})^{1/2}$ with $\sqrt{s}$ center of mass energy in proton-proton collision. $S^2(W^{\pm})$ are rapidity gap survival factor giving the probability that the rapidity gap is not populated by additional soft interactions involving the initial state spectators \cite{Khoze:2002dc,Jones:2013pga}. For the numerical calculations, we have used the values of  rapidity gap survival probabilities $S^2(W)$ from Ref. \cite{Jones:2016icr}.

The term $\mathrm{d} \mathrm{n} \over {\mathrm{d} \mathrm{k}}$ represents the photon flux \cite{Bertulani:2005ru,Kepka:2009zz}, expressed as 
\begin{eqnarray}
\frac {\mathrm{d} \mathrm{n}}{\mathrm{d} \mathrm{k}} =\frac{\alpha_{em}}{2\pi k}\Big[1+\Big(1-\frac{2k}{\sqrt{s}}\Big)^2\Big]\Big(
\ln \Omega-\frac{11}{6}+\frac{3}{\Omega}-\frac{3}{2\Omega^2}+\frac{1}{3\Omega^3}\Big),
\end{eqnarray}
where $\Omega=1+0.71/Q^2_{min}$, with $Q^2_{min}=k^2/\gamma^2_L$, $\gamma_L$ is the Lorentz boost factor with $\gamma_L=\sqrt{s}/2M_p$, $M_p$ is proton mass.

\begin{figure}[htbp]
\centering 
\includegraphics[width=8cm,height=8cm]{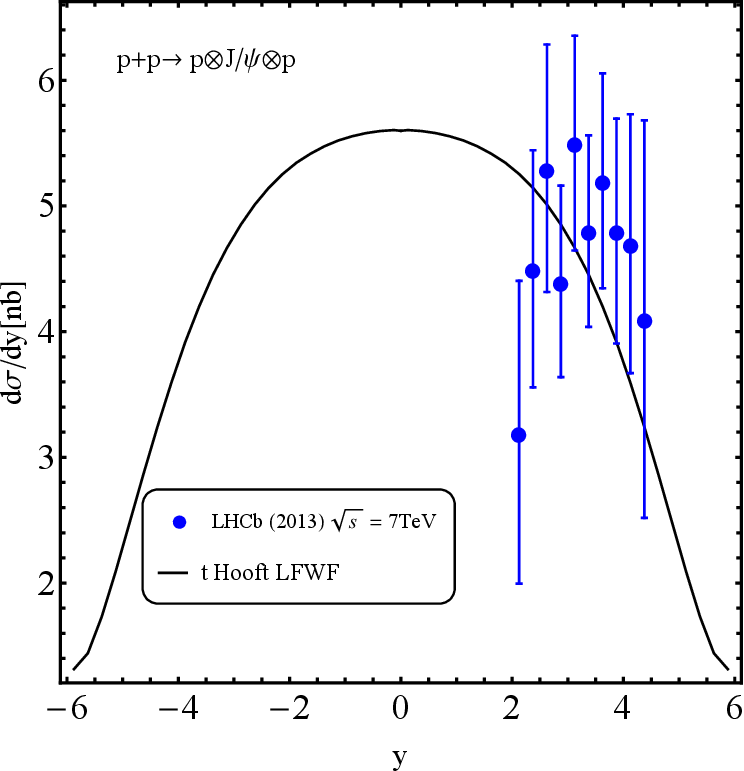}
\includegraphics[width=8cm,height=8cm]{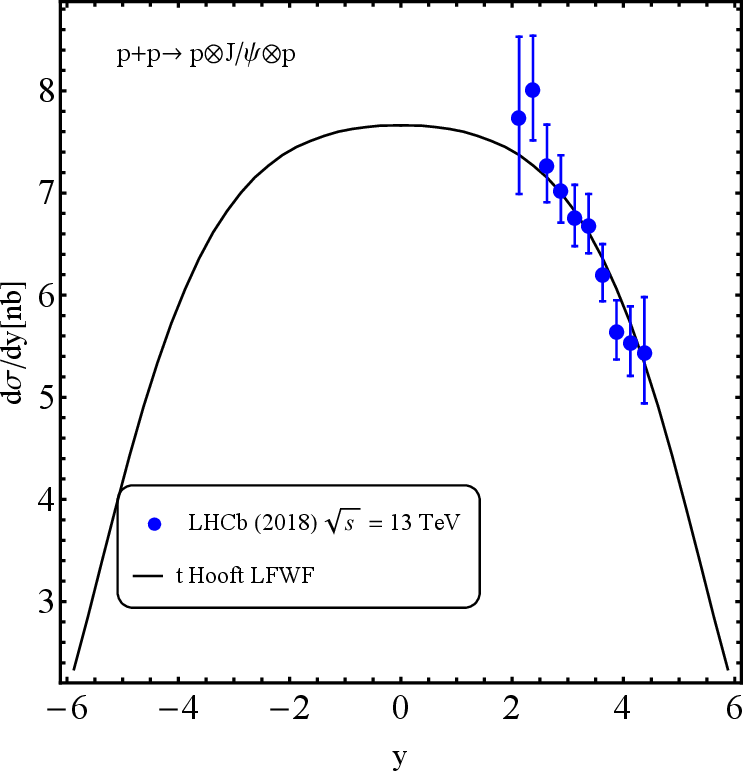}
\caption{Predictions for Differential scattering cross section as a function of rapidity $y$ for the exclusive $\mathrm{J/}\psi$ photoproduction off proton in ultraperipheral collision $\mathrm{p}$-$\mathrm{p}$  collisions  at $\sqrt s = 7 ~\mbox{TeV}$  compared to LHCb data \cite{LHCb:2013nqs,LHCb:2014acg}  and $\sqrt s = 13 ~\mbox{TeV}$    compared to LHCb data \cite{LHCb:2018rcm}}.
\label{upc:ppjpsi}
\end{figure}

\begin{figure}[htbp]
\centering 
\includegraphics[width=8cm,height=8cm]{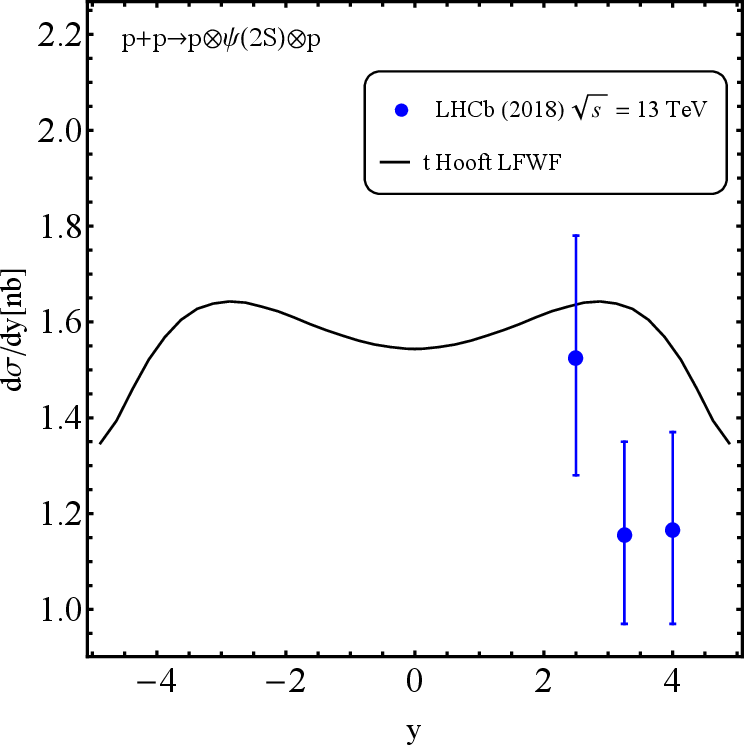}
\caption{Predictions for differential scattering cross section as a function of rapidity $y$ for the exclusive $\psi(2S)$ photoproduction off proton in ultraperipheral collision $\mathrm{p}$-$\mathrm{p}$  collisions  at $\sqrt s = 7 ~\mbox{TeV}$  compared to LHCb data \cite{LHCb:2013nqs,LHCb:2014acg}  and $\sqrt s = 13~ \mbox{TeV}$    compared to LHCb data \cite{LHCb:2018rcm}}.
\label{upc:pppsi}
\end{figure}

We will now  compare our theoretical predictions with the rapidity differential cross section data from LHCb for the process ${\mathrm p} {\mathrm p}  \to {\mathrm p}  \otimes J/\psi \otimes {\mathrm p} $. In Fig. \ref{upc:ppjpsi}, we presented the plots for differential scattering cross section in the dipole model as a function of rapidity ${\mathrm y} $ for the exclusive $\mathrm{J/}\psi$ photoproduction off proton in $p-p$  collisions  at $\sqrt s = 7~ \mbox{TeV}$   and $\sqrt s = 13 ~\mbox{TeV}$    compared to LHCb data \cite{LHCb:2013nqs,LHCb:2014acg}.  For exclusive $\psi(2S)$ photoproduction, we presented the plots for differential scattering cross section as a function of rapidity ${\mathrm y} $  at $\sqrt s = 13 ~ \mbox{TeV}$  compared to LHCb data \cite{LHCb:2018rcm} in Fig. \ref{upc:pppsi}. The agreement between our predictions and LHCb data is significant.

\begin{figure}[htbp]
\centering 
\includegraphics[width=8cm,height=8cm]{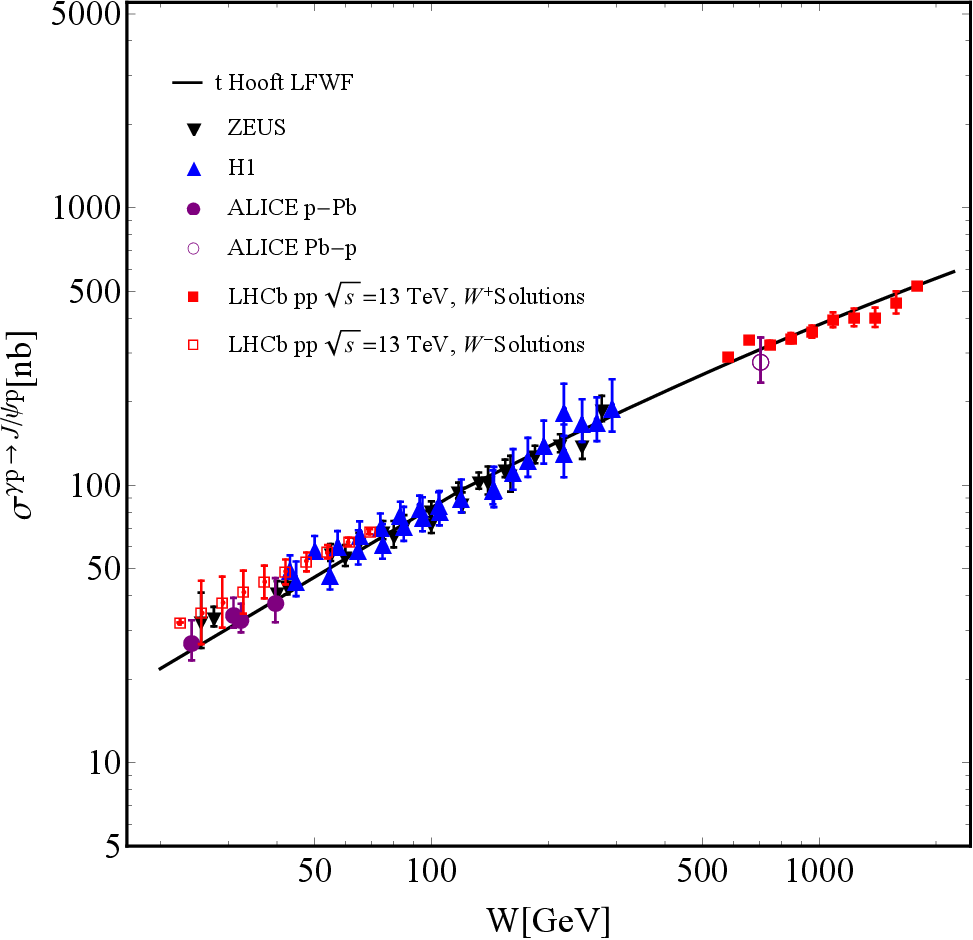}
\includegraphics[width=8cm,height=8cm]{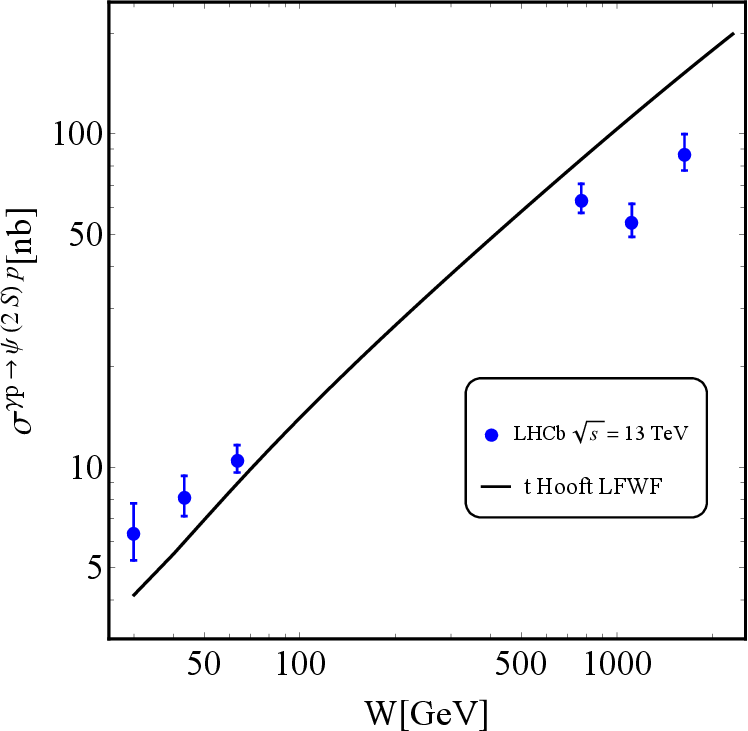}
\caption{Predictions for the $\mathrm{J/}\psi$ and $\psi(2S)$ photoproduction cross section as a function of the centre-of-mass energy of the photon-proton system. The  solid black line represent our prediction for the using the 't Hooft  LFWF. Results from the $ep$ collisions at  H1  and ZEUS Collaboration \cite{H1:2013okq,H1:2003ksk,H1:2005dtp,ZEUS:2002wfj,H1:2002yab}, $\mathrm{p}$-$\mathrm{Pb}$  collisions at $\sqrt{s_{NN}}=5.02 ~\mbox{TeV}$ ALICE Collaboration \cite{ALICE:2014eof}  and LHCb  Collaboartion at $\sqrt{s}=7, 13 ~\mbox{TeV}$ \cite{LHCb:2013nqs,LHCb:2014acg,LHCb:2018rcm} are also shown. }.
\label{jpsi:photo}
\end{figure}

The study of $\mathrm{J/}\psi$ photoproduction is important as  $\mathrm{J/}\psi$ photoproduction cross section  shed light on the small-$x$ gluon parton distributions of the target proton. In Fig. \ref{jpsi:photo}, we predicted the variation of photoproduction cross sections for $\gamma p \to \gamma J/\psi$ with $W_{\gamma p}$ with photon-proton center of mass energy. In case of photoproduction, the initial state photon is real and  only the transverse component of the overlap wavefunction contributes to the cross section.
We have compared our predictions to the H1 \cite{H1:2013okq,H1:2003ksk,H1:2005dtp,H1:2002yab}, ZEUS \cite{ ZEUS:2002wfj} and ALICE \cite{ALICE:2012yye} data. We have also presented a comparison to the LHCb results at $\sqrt s = 7~ \mbox{TeV}$ \cite{LHCb:2014acg} and  $\sqrt s = 13 ~\mbox{TeV}$ \cite{LHCb:2018rcm}. The recent results at $ 13 ~\mbox{TeV}$ are important as they extends the $W$ range to reach to almost 2 $\mbox{TeV}$.  
The experimental data labeled as ZEUS, H1 and ALICE are measured direct results  by these Collaborations whereas the LHCb results  at $\sqrt s = 13 ~\mbox{TeV}$  are  extracted using improved photon flux and rapidity gap survival factors in proton-proton UPCs \cite{LHCb:2018rcm}. The cross sections of LHCb measurements are divided by the rapidity gap survival factor and photon  flux as presented in Eq. \ref{dsdy}.

\begin{figure}[htbp]
\centering 
\includegraphics[width=8cm,height=8cm]{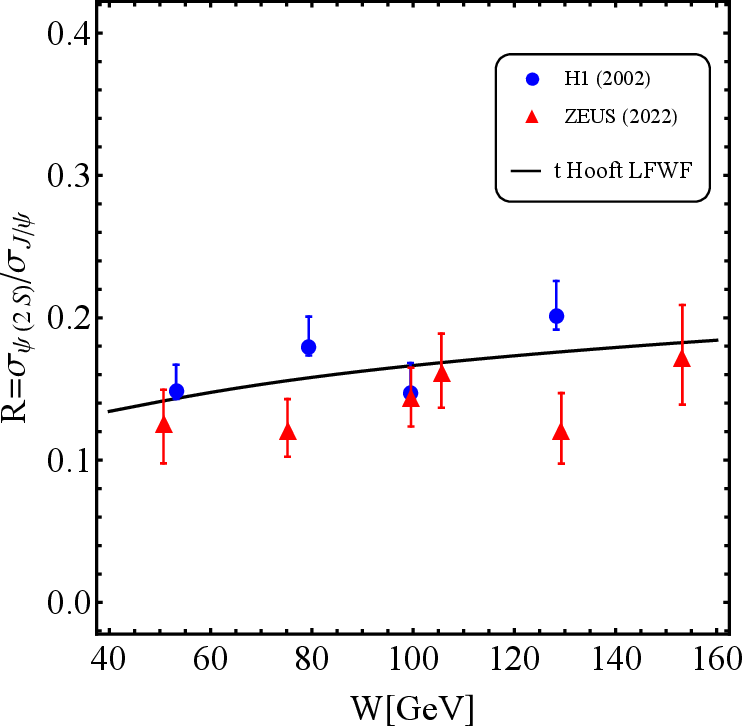}
\caption{ Predictions for the  ratio of photoproduction cross section $\psi(2s)$ to the $\mathrm{J/}\psi $ as a function of center of mass energy of the photon-proton system.
\cite{H1:2002yab,ZEUS:2022sxn} }.
\label{upc:ppR}
\end{figure}

Finally, we presented the cross section ratio $R= \sigma_{\psi(2S)} / \sigma_{\mathrm{J/}\psi}$ in exclusive photoproduction as measured in $ep$ collisions at HERA.  The cross section ratio was determined as a function of $W$ and plots are presented in Figure  \ref{upc:ppR}. Our model calculations were compared to the measured values of $W$ in H1 and ZEUS detector at HERA and give a reasonable description of the data. It is observed that ratio increases very slowly with increasing $W$ and the value of $R$ is almost constant. It is also important to mention that the $R$ value is very small signifying that $\psi(2S)$  cross section is much suppressed relative to the $\mathrm{J/}\psi$ cross section. This suppression is expected as the  $\psi(2S)$  wave function has a radial node close to the typical transverse separation of the virtual $c \bar c$ pair. 

In view of the above results, we conclude  that dipole model in conjunction with 't Hooft light wavefunctions with minimum set of free parameters give  the simultaneous description of the heavy vector meson $\mathrm{J/}\psi$ and $\psi(2S)$ diffractive and inclusive production data.

\section{Conclusion}
\label{Conclusion}

We have used the  dipole model together with the 't Hooft light-front vector meson wavefunction (LFWF) to compute the cross sections for diffractive $\mathrm{J/}\psi$ and $\psi(2S)$ mesons production. The parameters of holographic light-front meson wavefunction  were obtained by fitting the exclusive electroproduction cross section data at HERA data in $ep$ collisions. Dipole model together with LFWF is able to give the simultaneously description of differential scattering cross section, electroproduction, proton charmed structure function $F^2_{c \bar c}$, and photoproduction data with holographic mass scale $\kappa = 1.41~\mbox{GeV}$  and $m_c = 1.27 ~\mbox{GeV}$. 

We have also extended our work to investigate the diffractive $\mathrm{J/}\psi$ and $\psi(2S)$ photoproduction off nucleons in ${\mathrm p} {\mathrm p}$ ulptraperipheral collisions at the LHCb. Using the dipole formalism with 't Hooft LFWF,  we investigated the rapidity dependence $\mathrm y$ for the differential scattering cross sections for the  centre-of-mass energies.  A comparison has been made with the  recent LHCb Collaboration data at $\sqrt s = 7 ~\mbox{TeV}$ and $\sqrt s = 13 ~\mbox{TeV}$. We have also investigated the the ratio of cross section for $R = {\sigma_{\psi(2S)} \over \sigma_{\mathrm{J/}\psi}}$ and compared with H1 and ZEUS data. Our result supported the small value of ratio as expected. We conclude that the dipole model with 't Hooft LFWF give a good description of the experimental data for $\mathrm{J/}\psi$ and $\psi(2S)$ UPCs and photoproduction data. It will be interesting to investigate the impact of the LFWFs on final state observables in proton-nucleus and nucleus-nucleus collisions in future as new data is expected from Large Hadron Collider (LHC) run 3.

\section{Acknowledgments}
 The work of NS is supported by the grant from Science and Engineering Research Board, Government of India.


\end{document}